\documentclass[12pt]{article}
\begin{document}
\begin{flushright}
UNILE-CBR1
\end{flushright}
\begin{center}
{\large \bf Borel Resummation of the Perturbative Free Energy
of Hot Yang-Mills Theory}
\end{center}

\vspace{0.5in}
\begin{center}
{Rajesh R. Parwani}

\vspace{0.5in}

{Departimento di Fisica, Universita' di Lecce,\\}
{and\\}
{Istituto Nazionale di Fisica Nucleare\\}
{Sezione di Lecce\\}
{Via Arnesano, 73100 Lecce, Italy.}

\vspace{0.25in}
\end{center}

\begin{abstract}

The divergent perturbative expansion of the free-energy density of thermal
$SU(3)$
gauge theory is resummed into a rapidly convergent series using a 
variational implementation of
the method of conformal mapping of the corresponding Borel series. The resummed
result differs significantly
from non-perturbative lattice simulations and the discrepancy is attributed to
the presence of a pole on
the positive axis of the Borel plane.
The position of that pole is determined numerically and the difference between
the lattice data and the
resummed series is related to a phenomenological bag `constant'. 


\end{abstract}

\vspace{0.5in}

It is generally believed that the collision of heavy-ions at sufficiently high
energies will lead to
the formation of a new phase of matter, the quark-gluon plasma, and experiments
to produce such a plasma
are underway at Brookhaven and CERN. As the effective coupling,
$\alpha$, of quantum chromodynamics (QCD) decreases with
an increase in energy, theorists have used perturbative methods
to study properties
of the plasma at high temperature ($T$).
For example, a completely analytical calculation of the free-energy density of
QCD to order
$\alpha^{5/2}$ was performed a few years ago
\cite{A,BN}. The { \it purely gluonic} contribution is given by

\begin{eqnarray}
{F \over F_{0} } &=& 1 -{15 \over 4} \left({\alpha \over \pi}\right) + 30
\left({\alpha \over \pi}\right)^{3/2}
+ \left( 67.5 \ln\left({\alpha \over \pi}\right) + 237.2 -
20.63\ln\left({\bar{\mu} \over 2\pi T}\right) \right) \left({\alpha \over
\pi}\right)^2
\nonumber \\
&& \, -\left(799.2 - 247.5\ln\left({\bar{\mu} \over 2\pi T}\right) \right)
\left({\alpha \over \pi}\right)^{5/2} \, ,
\label{ym}
\end{eqnarray}
where $F_{0} = {-8\pi^2T^4 \over 45}$  is the contribution of non-interacting
gluons and $\bar{\mu}$ is the
renormalization scale in the $\overline{MS}$ scheme. Unfortunately (\ref{ym})
is an oscillatory, non-convergent, series even for $\alpha$
as small as $0.2$, which is close to the value of physical interest.

Pade$'$ Approximants (P.A.'s) were used in Ref.\cite{HK} to resum
the series (\ref{ym}).
It was found that the dependence on the arbitrary scale $\bar{\mu}$
was reduced and the convergence of the series somewhat
improved. However P.A.'s have a number of well-known drawbacks. 
For those and other reasons, in Refs.\cite{BB}
the authors abandoned the expansion of the
free-energy density with respect to the coupling constant and considered instead
selective resummations of gauge-invariant diagrams.
Though the results of \cite{BB}
compare favorably with lattice simulations \cite{latt},
calculations beyond leading order
are complicated and thus it seems that the issue of convergence is left open.
In Ref.\cite{kaj}, yet another procedure was used to study
the free-energy of hot QCD: short distance perturbative effects were handled
analytically while long-distance effects were described \cite{BN}
by an effective
three-dimensional theory and studied numerically.

It appears that a number of separate issues concerning Eq.(\ref{ym}) have become
confused in the literature.
The first issue is whether the given divergent series can be resummed into a
convergent series, preferably in a
systematic and well-motivated way.
The second issue is whether such a resummed series accurately represents the
physical quantity. The third issue is whether one obtains any new physical
insight
in the process. It will be the attempt of this paper to shed some light on these
and
related questions.

Recall that the divergence of perturbative expansions in quantum field theory is
is a generic phenomena \cite{Z}. Given a series

\begin{equation}
\hat{S}_{N}(\lambda) = 1 + \sum_{n=1} ^{N}  f_{n} \lambda^{n} \, ,
\label{ser}
\end{equation}
where $\lambda$ is the coupling constant, one expects the coefficients $f_n$
to grow as $n!$ for large $n \ $.
It is natural then to introduce the Borel transform

\begin{equation}
B(z) = 1 + \sum_{n=1}^{N} {f_{n} z^{n} \over  n!}
\label{borel}\end{equation}
which has better convergence properties than (\ref{ser}). The series (\ref{ser})
may
be recovered from (\ref{borel}) through a Laplace transform

\begin{equation}
\hat{S}_{N}(\lambda) = {1 \over \lambda} \int_{0}^{\infty} \ dz \ e^{-z/\lambda}
\ B(z) \, .
\label{lap}
\end{equation}

In order to proceed non-trivially, one first performs an approximate
summation of the series (\ref{borel}) so that Eq.(\ref{lap}) then gives the
resummed version of
(\ref{ser}).
Now, suppose that the only singularity of $B(z)$ is at
$z=-1/p$, with $p$ real and positive.
Then the radius
of convergence of the Borel series is $1/p$.
In order to perform the integral in
(\ref{lap}), one needs to extend the domain of
convergence of the Borel series.
One way to do this is by the method of conformal mapping\cite{Z,GZ}.
Define

\begin{equation}
w(z) = { \sqrt{1 + pz} -1 \over \sqrt{1+pz} +1 }
\label{con}
\end{equation}
which maps the Borel plane to a unit circle. The inverse of (\ref{con}) is given
by

\begin{equation}
z= { 4w \over p} {1 \over (1-w)^2 }
\label{inv}
\end{equation}

The idea is to rewrite (\ref{borel}) in terms of the variable $w$. Therefore,
using (\ref{inv}),
$z^n$ is expanded to order $N$ in $w$ and substituted into
(\ref{borel},\ref{lap}). The result
is

\begin{equation}
S_{N}(\lambda) = 1 + {1 \over \lambda} \sum_{n=1}^{N} {f_{n} \over n!} \left({ 4
\over p}\right)^n \, \sum_{k=0}^{N-n}
{ ( 2n+k-1)! \over  k! \ (2n-1)!}
 \int_{0}^{\infty} e^{-z/\lambda} \ w(z)^{(k+n)} \, dz \; ,
\label{rser}
\end{equation}
where $w(z)$ is given by (\ref{con}). Equation (\ref{rser}) represents
a resummation of the original series
(\ref{ser}). This technique has been used in determining critical exponents
in statistical systems where the singularity
at $z=-1/p$ is due to instantons \cite{Z,GZ}, and in QCD at zero temperature
where the singularity is due
to renormalons \cite{qcd}.

Currently no information is available about the exact location of singularities
in
the Borel plane of thermal QCD though undoubtedly there is at least one on the
negative semi-axis. Therefore in order to apply the resummation
(\ref{rser}) to Eq.(\ref{ym}), a new idea is introduced
in this paper: It is first assumed that the only singularity is at
$z=-1/p,\ p>0$,
with the value of $p$ determined by the condition that it be
the position of an extremum of
(\ref{rser}). That is, $p$ is chosen to be a solution of

\begin{equation}
\left({\partial S_{N}(\lambda,p) \over \partial p} \right)_{\lambda=\lambda_0}
=0 \, .
\label{ext}
\end{equation}

Since $S_{N}(\lambda,p)$ depends on the coupling $\lambda$, one first fixes
$\lambda$ at some reference
value $\lambda = \lambda_{0}$ (say, at the mid-point
of the range of interest) in order to solve Eq.(\ref{ext}). Fortunately, it
turns out
that the solution of (\ref{ext}), and hence the convergence of $S_N$,
is not very sensitive to the exact value of $\lambda_{0}$.

Let me illustrate the technique by applying it to two cases where exact results
are known.
Consider first the integral
\begin{equation}
I(\lambda) = \int_{0}^{\infty} \ dz \, {e^{-z}  \over 1+z\lambda} \; .
\label{toy}
\end{equation}
If the right-hand-side of (\ref{toy}) is expanded as a power series in $\lambda$
one obtains at $N$-th order

\begin{equation}
I_{N}(\lambda) = 1 + \sum_{n=1}^{N} \lambda^n (-1)^n n! \; .
\label{toyser}
\end{equation}
Clearly, from (\ref{toy}), the exact Borel transform of this series is $B(z) =
1/(1 +z)$ with a singularity
at $z=-1$. Ignoring this information, let us resum the divergent series
(\ref{toyser})
using (\ref{rser}) with $f_n = (-1)^n n!$ and values of $p$ determined for each
$N$ through equation
(\ref{ext}) at the reference value $\lambda_{0}=0.5$. The results for
(\ref{ext})
are as follows:
There is no extremum for $N=1$. For $N=2$,
there is a minimum at $p=2.65$. For $N=3$ there is a local maximum at $p=1.5$
and a
minimum at $p=5.1$. For $N=4$ there is a local minimum at $p=1.3$, a local
maximum at $p=2.3$
and a global minimum at $p=8.4$. Jumping ahead to $N=8$, there is a local
minimum at $p=1.075$, a local maximum at $p=1.5$
and a global minimum at $p=3.6$.

Thus in general (\ref{ext}) has more than one solution for a given $N$ and
$\lambda_{0}$.
In Fig.(1), Eq.(\ref{rser}) is plotted for $N=2,3,4$ and $8$, at the
respective minimum. Notice
how the curves converge rapidly to the exact value given by (\ref{toy}).
Alternatively,
one might choose for each $N$,
the value of $p$ (from the multiple solutions of
Eq.(\ref{ext})) which seems to be part of a converging sequence. In this case
the values,
$p=2.65 (N=2),
p=1.5 (N=3), p=1.3 (N=4)$, and $p=1.075 (N=8)$, appear to converge to the exact
value
$p=1$. The curves are shown in Fig.(2). Clearly the curves in Fig.(2) converge
faster to the exact value than those of Fig.(1), but unless one is interested in
very high numerical accuracy,
the difference is not significant. For example, at $\lambda=0.5$, the exact
value
of (\ref{toy}) is $0.722657$, while the resummed value for $N=8$ is given at the
global minimum $p=1.075$
by $S_{8}(\lambda=0.5,p=1.075) = 0.722652$, and at the local minimum $p=3.6$ by
$S_{8}(\lambda=0.5,p=3.6) =0.722524$.

The main points illustrated by this example, which seem to be common to the
other cases studied,
are, (i) the rapid convergence of the resummed series represented by
(\ref{rser}) compared to the original
wildly oscillating series (\ref{toyser}), (ii) the relative insensitivity of the
convergence and numerical value
of the resummed series to the particular extremum chosen among the
possible multiple solutions of (\ref{ext})
 (for a given $N$ and $\lambda_{0}$), even if the chosen value of $p$ is quite
different from the
exact value, (iii) the relative insensitivity of Eq.(\ref{ext}),
and hence the convergence of (\ref{rser})
to the precise value of $\lambda_{0}$.

For another example, consider thermal $O(M)$ $\lambda^2 \phi^{4}_{4}$
field theory
in the limit $M \to \infty$. The exact free-energy density in this case
has been determined in Ref.\cite{L}. In Eq.(5.8) of
that paper\footnote{Note that the definition of the coupling constant used here
is
different from that in Ref.\cite{L}.} the perturbative
expansion, in $\lambda$, of the free-energy density is also given up to
$\lambda^{6}$.
Defining $S=(F(T)-F(0))/F_{ideal}$
and choosing $\overline{\mu}=T$ for simplicity, the values of
$f_n$ for $2 \leq n \leq 6$ can be read off from Eq.(5.8)  of Ref.\cite{L}
and (\ref{ext}) solved at some  reference value, say $\lambda_{0}=4$.
 The solutions of
(\ref{ext}) in this case are: $N=3, p=0.1$(min); $N=4,p=0.05$(local max),
$p=0.2$(min); $N=5, p=0.025$ (local min), $p=0.1$ (local max),
$p=0.3$(global min); $N=6, p=0.1$ (local max), $p=0.45$ (global min).
As in the first example, the convergence of the resummed series is found to be
rapid even for large coupling,
in contrast to the oscillatory behaviour of the
ordinary pertubation expansion observed in \cite{L}.
Fig.(3) shows the curves for
$S_{N}(\lambda), 3 \leq N \leq 6$, for the value $p=0.1$, which seems to be the
value
$p$ converges to as $N$ increases.
The exact value of the free-energy density
at $\lambda=8$ taken from Fig.(6) of Ref.\cite{L} is about $0.875$. By
comparison
the resummed value predicted here is given at sixth order by
$S_{6}(\lambda=8,p=0.1)
=0.889$. On the other hand, if one evaluates
$S_{6}$ not at $p=0.1$
but rather at its global minimum $p=0.45$, one gets
$S_{6}(\lambda=8,p=0.45)=0.849$, a difference of less than $5\%$.

Actually, no information is available about the singularities in the Borel plane
for the $O(M \to \infty)$ scalar field studied in \cite{L}.
The good agreement of the results obtained here with the exact results
of Ref.\cite{L} leads one to conjecture that for the free-energy of this model,
the singularity closest to the origin in the Borel plane might be near $p=0.1$,
that
is, $z=-10$.

Note that although the coupling constant in the scalar
field theory model is $\lambda^2$, the perturbative
expansion  of the free-energy density contains the odd powers $\lambda^3$ and
$\lambda^5$ which is
typical of thermal
field theories with massless particles (or at very high temperatures)
\cite{A,K}.
Physically these
are due to collective effects such as Debye screening and it is sometimes
suggested
in the literature
that these terms should be treated on a different footing. However, as the
analysis
above shows, from a mathematical point of
view these odd powers are no different from the other terms in the
expansion of the free-energy density and can be resummed as part of
a single series.

Finally, the resummation technique of Eqs.(\ref{rser},\ref{ext}) is applied to
the free-energy density of $SU(3)$
gauge theory given in Eq.(\ref{ym}).
As in Refs.\cite{BB}, I replace $({\alpha \over \pi})^{1 \over 2}$ by the
approximate two-loop running coupling constant defined by
\begin{equation}
\lambda(c,x) = {2 \over \sqrt{11 L(c,x)}} \left( 1-{51 \over 121}
{\ln(L(c,x)) \over L(c,x)} \right)
\label{run}
\end{equation}
where $L(c,x) = \ln((2.28 \pi c x )^2)$, $c=\bar{\mu} / 2 \pi T$ and $x=T/T_c$,
with $T_c \sim 270 MeV$ the critical temperature which separates the low and
high temperature phases \cite{latt}. Furthermore, as in Ref.\cite{HK}, I have
absorbed the
$\ln(\alpha)$ term which appears at three-loop order into the coefficient
of the $\alpha^2$ term in (\ref{ym}).

Fixing first the reference values $c_{0}=1,x_{0}=3$
(which fixes the reference value of $\lambda_{0}$),
the results of (\ref{ext}) are: $N=2$, no extremum; $N=3,p=3.2$(min); $N=4,
p=7.6$(min); $N=5,p=13.1$(min).
Since $c$ and $x$ appear in (\ref{ym}) and (\ref{run}) only logarithmically,
changing these
values in the range of say, $0.5< c < 2$, $\ 2<x<5$, has almost no impact on the
solution of Eq.(\ref{ext}) and hence on the optimal values of $p$.

The curves for the resummed series are plotted in Fig.(4) for $c=1$, that is at
the
renormalization scale $\bar{\mu}=2 \pi T$. Again the rapid and monotonic
convergence is manifest, the result
approaching the ideal gas value even at moderate temperatures $\sim 2 T_c$.
One can estimate the effect of the unknown higher order, $\lambda^6$,
contribution. It turns out to be negligible \cite{rp2}.
Therefore one feels confident that the $N=5$ curve in Fig.(4) is numerically close
to the (unknown) exact sum of the {\it perturbation series}.
In Fig.(5) the curve for $S_{5}(x,c,p=13.1)$ is plotted for three values of $c$
to indicate
its mild dependence (less than $1 \%$) to the arbitrary renormalization scale
$\bar{\mu}$.

Lattice results for the free-energy
density of pure $SU(3)$ theory are shown in Fig(6). The lattice community has
indicated
that their errors are under control (less than  $5 \%$). In that case,
I am left with the task of
explaining the significant difference (e.g. $\sim 15 \%$ at $T=3T_c$)
between the best analytically resummed result
represented by the $N=5$ curve in Fig.(4) and the lattice data.
At zero temperature, it is known that non-abelian gauge theories are not Borel
summable
\cite{qcd}.
That is, $B(z)$ contains singularities for positive $z$, rendering the integral
in
(\ref{lap}) ambiguous. The situation is not expected to be different at non-zero
temperature.
Usually \cite{qcd}, the presence of such singularities is taken to indicate the
existence of non-perturbative corrections.
One can estimate the ambiguity, $\delta S$, and hence the non-perturbative
correction,
as the residue of the integrand in (\ref{lap})
at the location of the singularity \cite{qcd}. If the singularity of
$B(z)$, on the positive semi-axis, closest to the origin is a pole at
$z=q$, then from (\ref{lap})

\begin{equation}
\delta S = {A \over \lambda} \ e^{- q/\lambda}
\label{bag}
\end{equation}
where $A$ is a constant. Assuming that the difference between the lattice data
and Fig.(5), is due to (\ref{bag}), the constants $A$ and $q$ can be determined
by
rewriting (\ref{bag}) as
\begin{equation}
\ln(\lambda \delta S) = \ln(A) -q/\lambda
\label{fit}
\end{equation}
and using for $\delta S$ the difference between Fig.(6) and the median curve in
Fig.(5)
(i.e. $c=1$).
Fig.(7) shows the left-hand-side of (\ref{fit}) plotted against $1/ \lambda$.
This
gives $A=e^{8.7}$ and $q=2.62$, and so, with $\lambda(x) \equiv \lambda(c=1,x)$,

\begin{equation}
S_{latt} = S_{pert} - {1 \over \lambda(x)} e^{8.7 - 2.62/\lambda(x)} \ ,
\label{bbag}
\end{equation}
where $S_{latt}$ represents the lattice data for the free-energy, and $S_{pert}$
the Borel resummed perturbative result, both normalized with respect to the
ideal gas value.

It is extremely reassuring that both the sign and magnitude of $q$ determined in
this way are self-consistent
with the assumptions made.
In particular, the singularity at $z=q=2.62$ is more than 30 times away from the
origin than the
singularity at $z=-1/p=-1/13.1$ and justifies { \it a posteriori} the
resummation
procedure (\ref{rser}) which considered only the nearest singularity.
Furthermore since $S_{pert}$ is extremely close to
the ideal gas value, Eq.(\ref{bbag}) may be interpreted as a generalisation of
phenomenological equations of state \cite{K} for the free-energy  where the
second term on the right-hand-side
of (\ref{bbag}) is called a 'bag constant'. In our case the `constant' is really
temperature dependent and represents a non-perturbative contribution to the
free-energy that
vanishes at infinite temperature. Note that this interpretation of the second
term on the
right-hand-side of (\ref{bbag}) is consistent with the usual one only because
the
resummed perturbative result lies {\it above} the lattice data.

Let me now summarize the main results of this paper. A new procedure, a
variational version of
the well-known conformal-mapping of Borel series, was introduced and
its practicability illustrated. It was shown that the badly divergent series for
the
free-energy density of thermal $SU(3)$ gauge theory (\ref{ym}) could be
resummed in  a systematic and
relatively well-motivated way into a rapidly convergent series. However the
final result differed
significantly from non-perturbative lattice data, suggesting
that the discrepancy is due to the non-Borel-summability of the theory.
Numerically, the difference was parametrised in terms of two constants (see
(\ref{bbag})),
and it is suggested that the ambiguity in the Borel integral is
the bag `constant' used in phenomenological models for the free-energy density.

In physical terms, the situation may be described as follows. {\it If} the Borel
resummed
perturbative expansion had agreed with the lattice data, then one
would have argued that the high-temperature phase of $SU(3)$ theory is
appropriately
described by weakly coupled gluons. However, if the
deviation of lattice data from the resummed and  convergent
 perturbation expansion found here is taken at face value, then one is
led to conclude that even at temperatures as high as
three times $T_c$ ($ \sim 700 MeV$), the phase of thermal $SU(3)$ is not
accurately
described by weakly coupled gluons but rather by more complicated structures
which are responsible for the non-perturbative bag contribution.
If one accepts this conclusion, then one must also be open to the possibility
that
what will be produced at CERN and Brookhaven  might be better characterised
as something other
than a quark-gluon plasma.  For some alternative descriptions of the
high-temperature phase of
QCD see, for example,  \cite{new}.

An extension of the analysis presented here to QCD and other 
thermal gauge theories,
a further development of the methodology itself and its 
applications to other physical problems, will be presented 
in an accompanying series of papers \cite{rp2}.

\vspace{0.3in}
{\bf Acknowledgements}: I thank Claudio Coriano$'$ for stimulating discussions
and
hospitality at Martignano, and the Department of Physics at Lecce for financial
support.
I also thank the YITP at Stony Brook for hospitality during the final stages of
this work.

\newpage

{\bf Figure Captions}

Figure 1: Plot of Eq.(\ref{rser}) for the model in Eq.(\ref{toyser}),
for $N=2,3,4$ and $8$ at the respective global minimum of Eq.(\ref{ext}).
The curves move upwards with increasing $N$. The curves for $N=3,4$
practically coincide, while  the
 curve for the exact
expression
(\ref{toy}) is indistinguisable from that for $N=8$.
\\

Figure 2:
Plot of Eq.(\ref{rser}) for the model in Eq.(\ref{toyser}),
for $N=2,3,4$ and $8$ at the respective values of $p$ from
Eq.(\ref{ext}) which converge to $1$. The values of the curves at $\lambda=0.5$
are, $N=2(0.704), N=3(0.726), N=4(0.7219), N=8(0.7227)$.
The curve for the exact expression
(\ref{toy}) is hardly distinguisable from that for $N=3,4,8$.
\\

Figure 3:
Plot of Eq.(\ref{rser}) for the model in Ref.\cite{L}
for $N=3,4,5$ and $N=6$ at the value $p=0.1$.
The curves move upwards as $N$ increases.
For $N=5,6$ they are practically identical.
The exact curve in Ref.\cite{L} lies slightly below
that for $N=6$: At  $\lambda=8$ the exact curve has the value
$0.875$.
\\

Figure 4:
The resummed (\ref{rser})
free-energy density of hot $SU(3)$ gauge theory
for $N=3,4$ and $5$, at the renormalization point $\bar{\mu} = 2 \pi T$.
The curves move upwards as $N$ increases.
\\

Figure 5:
The fifth order resummed free-energy density (\ref{rser}) of hot
$SU(3)$ gauge theory at $p=13.1$, for
the renormalization scale values $\bar{\mu}=0.5,1$ and $2$. The
free-energy density increases with increasing $\bar{\mu}$.
\\

Figure 6:
Mean lattice results for the free-energy density of hot $SU(3)$ gauge
theory from Ref.\cite{latt}. Here $S_{latt}$ refers to the free-energy
divided by the free-energy of an ideal gas of gluons.
\\

Figure 7:
Plot of the left-hand-side of Eq.(\ref{fit}) against $1 / \lambda$ at
several points (temperatures).

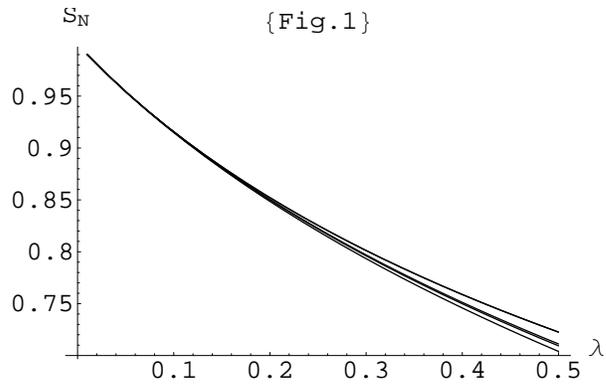
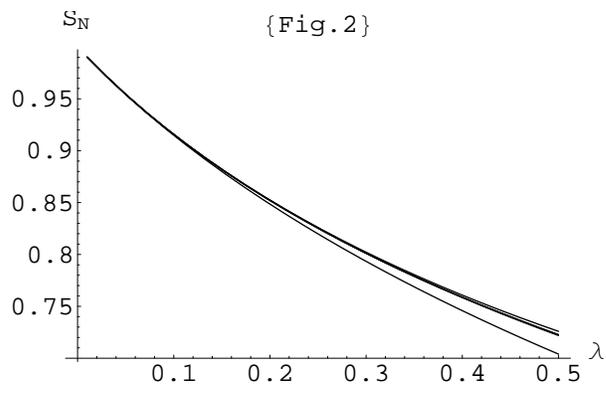
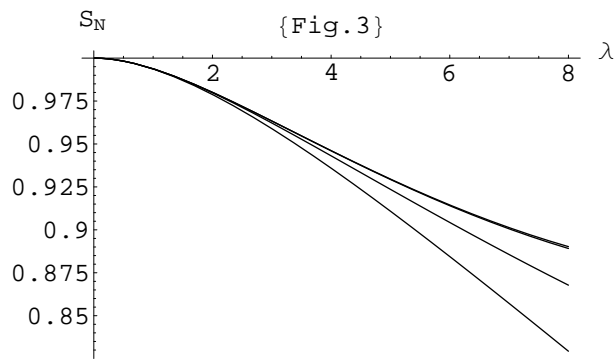
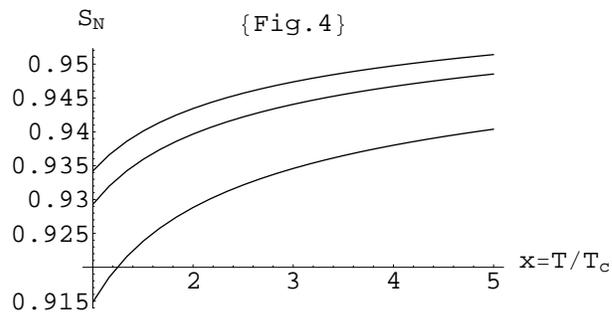
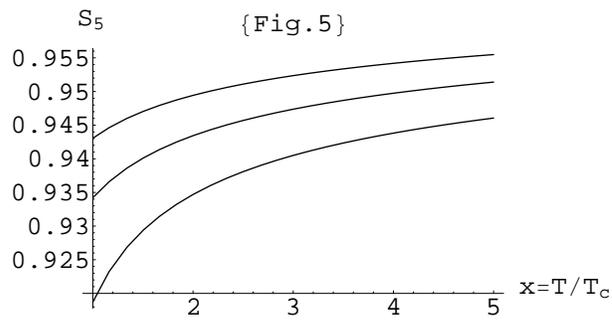
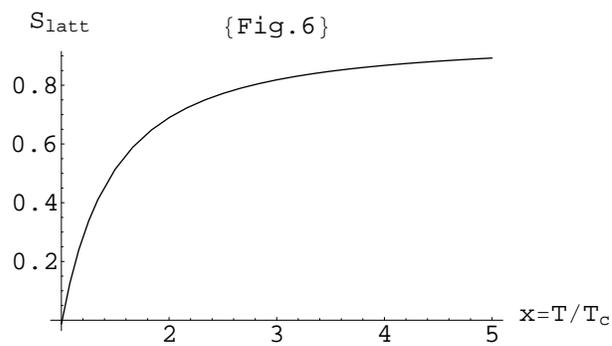
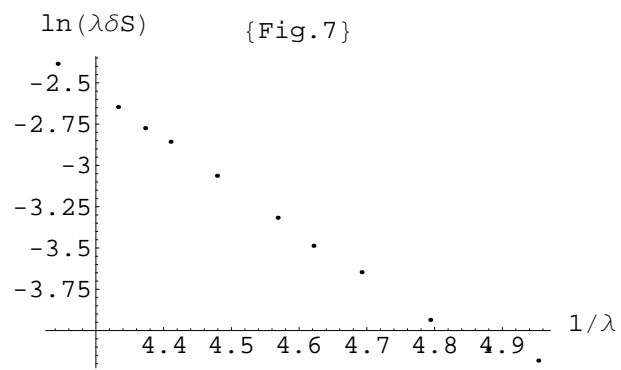
\end{document}